# Author's Post-Print (Final draft post-refereeing)

# Hole Detection in WSN with Force-directed Algorithm and Transfer Learning

**Yue-Hui Lai** · **Se-Hang Cheong** · **Hui Zhang** · **Yain-Whar Si**

**Abstract** Hole detection is an important research problem in wireless sensor network. However, distributed approaches proposed in recent years for hole detection problem have high computational complexity. In this paper, we propose a novel approach for hole detection in wireless sensor network called FD-TL which is based on layout generation capability of Force-directed Algorithms and image recognition power of Convolutional Neural Network with transfer learning. In contrast to existing approaches, the proposed approach is a pure topology-based approach since FD-TL can detect both triangular and non-triangular holes from a wireless sensor network based on the input network topology without relying on the physical locations of the anchor nodes. In FD-TL, a Force-directed Algorithm is used to generate a series of possible layouts from a given input topology. Next, a Convolutional Neural Network is used to recognize potential holes from the generated layouts. During the training phase, a transfer learning method is used to aid the recognition process. Experimental results show that FD-TL method can achieve 90% sensitivity and 96% specificity for hole detection in wireless sensor networks.

Yue-Hui Lai
University of Macau, Macau, China
E-mail: mb95427@um.edu.mo

Se-Hang Cheong
University of Macau, Macau, China
E-mail: dit.dhc@lostcity-studio.com

Hui Zhang
United International College, BNU-HKBU, Zhuhai, China
E-mail: amyzhang@uic.edu.cn

Yain-Whar Si
University of Macau, Macau, China
Tel.: +853 8822-4454
E-mail: fstasp@umac.mo

# 1 Introduction

In recent years, wireless sensor networks (WSN) are increasingly used in various applications involving military aviation [51], residential properties [57], and health care [36]. A WSN is composed of multiple sensors with perceptive ability and the aim of each sensor is to collect and process the perceived objects' information [52]. In WSNs, data is transmitted directly from one sensor to another which are located within the communication range. When the battery power of a sensor is exhausted or the distance between the sensor and its neighbours exceeds the communication range, a collection of such sensors could form a coverage hole within a WSN. Due to these coverage holes, a WSN may not be able to accurately collect information of sensors in the affected area. Therefore, it is crucial to detect and rectify coverage holes [58].

Previous work on hole detection in WSNs can be categorized as geometrical, statistical, and topological methods [50]. In geometrical methods [35][56][44], extensive computation is performed at the nodes for estimating the holes. Before the computation, the position of the nodes needs to be known in advance. Next the critical nodes are computed in geometric structure such as Voronoi diagram and Delaunay triangulations to find the coverage holes. Due to the requirement of a huge amount of computations in geometrical methods, complexity of these approaches can be high. In contrast to geometrical approaches, statistical hole detection algorithms assume that the WSNs contain dense distribution of sensor nodes [14][13] or random distribution of sensor nodes [31]. Therefore, statistical methods are limited to wireless sensor networks with high-density or random node distribution. In topological methods, the sensor's exact geographic information is not required for the hole detection. However, the topological attributes(connectivity information) among the nodes is needed for detecting holes. Some topological algorithms [10][45][53] are mainly based on homology computation [55], which cannot effectively detect triangular coverage holes.

In order to alleviate the limitation of existing hole detection algorithms, we propose a novel approach called FD-TL which only utilizes the input network topology. Specifically, FD-TL only requires the information about the unique identification of nodes (node ID) and the list of edges connecting the nodes in a WSN. In contrast to existing approaches, FD-TL does not use physical location of nodes during the hole detection. The proposed FD-TL method is a centralized approach and can be used to detect both triangular and non-triangular holes effectively. Compared with distributed approach, the proposed FD-TL method has no need to make complex computaion for sensor nodes.

FD-TL is based on Force-directed (FD) algorithm and Convolutional Neural Network (CNN). FD algorithms are flexible, intuitive and easy to implement. FD algorithms can be used to generate high-quality layouts of the input typologies without the need of node location information [5]. CNNs were first developed in the 1980s. CNNs had not attracted broad attention and were not widely used until 2012. In 2012, AlexNet [24] broke the image classification record in ImageNet challenge [41]. Since then, CNNs are increasingly used in various applications.

The proposed method does not depend on the location information generated by GPS or anchor nodes. FD-TL is based entirely on network topology information and therefore sensors of a WSN can be distributed randomly. The hole detection process of proposed FD-TL method is shown in Figure 1. In FD-TL, a possible layout of a wireless sensor network is generated by the FD algorithm from a given input network topology. A generated layout can be considered as a potential sensor distribution (estimation) among hundreds of such possibilities. Next, a trained CNN model is used to detect holes from the layout and output the detection results.

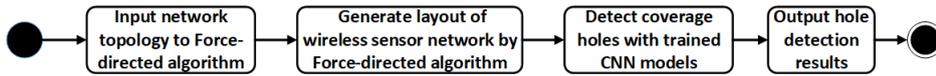

Fig. 1: Coverage hole detection process with FD-TL method.

Specifically, the CNN model performs two tasks in the FD-TL: detecting the position of the holes and segmenting the hole's boundary to identify the identity of the nodes along the border of the hole. In this paper, we use Mask R-CNN [16], Mask R-CNN with FPN [29], TensorMask [4], PointRend [23] and BlendMask [3] respectively for hole detection. In FD-TL method, we train CNN models by using the transfer learning method [37]. Transfer learning method is widely in image recognition and computer vision applications when the training data is insufficient. Transfer learning can also be used to save training time and computing resources. In FD-TL, pre-trained CNN models of Microsoft COCO data set [30] are fine-turned with our own hole detection training data set. Since there is no coverage hole data set available in the public domain for WSNs, we create a labelled data set for training the CNN models from scratch.

The main contributions of our approach are as follows:

– A novel method called FD-TL for hole detection in WSN is proposed, which can achieve higher sensitivity and specificity for hole detection in WSNs.
– Transfer learning method was used to alleviate the problem of insufficient training data.
– Our trained CNN models can detect holes from a generated layout within 1 second.

The rest of the paper is organized as follows. In Section 2, we review the related work. In Section 3, we introduce the overview of proposed FD-TL method. In Section 4, hole detection process by FD-TL is detailed. In Section 5, experiments are conducted to evaluate to effectiveness of the proposed approach. In Section 6, we conclude our work with future work.

## 2 Related Work

In WSNs, all sensors have similar status [1]. These sensors are often equipped with battery power to operate. However, the lifespan of the batteries are limited. Since replacement cannot be immediately carried out in many application scenarios, the effected sensors may no longer function in these networks. Such situations can lead to the formation of coverage holes in WSNs.

Coverage holes can be also caused by external factors such as earthquake or thunderstorm. In these situations, some of the sensors could be entirely destroyed or malfunctioned. Coverage holes formed during such situations may affect the sensors' role in monitoring and transmitting of collected data. Therefore, detecting coverage holes is one of the important research problems in WSNs [58].

Existing coverage hole detection methods can be roughly categorized as geometrical, statistical, and topological methods respectively [50]. Geometrical methods utilize the geographic coordinates of the sensor [55]. Topological methods utilize the connectivity information of the network. However, these methods do not require exact node location information for detection [55]. In statistical method, the distribution of sensors follows certain statistical functions [2].

2.1 Geometrical methods

Geometrical methods assume that the location of the sensors and the relative distance between the sensors can be obtained by using Global Positioning System (GPS) or other positioning devices. In these methods, geometric figures such as Voronoi diagram and Delaunay triangulations are used to identify coverage holes with location information of the nodes [55].

In [35], Meguerdichian et al. proposed a centralized computation model. By combining the computational geometry and graph theoretic techniques such as Voronoi diagram, an optimal polynomial time algorithm was established for coverage calculation. In [43], So and Ye used Voronoi diagrams in a centralized manner to solve coverage problem in WSNs. Each node was used to detect the vertices of the Voronoi cells. Next the Voronoi diagram was used for coverage hole detection. Zhang et al. [56] proposed a kind of Voronoi diagram-based screening strategy to detect holes. In their work, Zhang et al. introduced a new virtual edge method to compute the boundary nodes. The information about the hole's position and shape was calculated in the proposed method. In [44], Soundarya et al. proposed a Delaunay-based method that incorporated the virtual edge, which could detect the coverage holes more accurately than existing tree-based methods [28]. In [34], Ma et al. proposed a computational geometry approach based on distributed hole detection protocol to detect the coverage holes in a post deployment scenario, where the sensing range of each node was uniform. However, all of these geometrical methods require node location information.

In summary, in geometrical methods, it is necessary to obtain the geographic information of the sensor to detect the coverage holes. When exact node location information is unavailable, the geometrical method cannot detect the coverage hole effectively. Therefore, geometrical methods are not suitable for hole detection when the geographic information is not available.

2.2 Statistical methods

Statistical methods for hole detection assume that sensor distribution follows some statistical functions [2]. In [14], Fekete et al. proposed an approach to identify the boundary nodesby applying methods from stochastics, topology and geometry. The statistical arguments were also used to provide threshold to differentiate boundary nodes. The proposed approach aimed to solve the hole detection problem of large-scale dense sensor network. It was developed for determining the region's topology, Voronoi diagram of boundaries and the outside boundary. However, their approach can only be applied to wireless sensor network with high-density node distribution. In [31], the optimal sensor mobility strategy and corresponding target mobility strategy were used to detect coverage holes. In their approach, Liu et al. assumed that the WSN contained a random distribution of sensor nodes. In [13], Fekete et al. proposed a distributed protocol. According to [13], the topological boundary information could be extracted by using restricted stress centrality [14]. However, the sensor networks considered in [13] contained large number of densely and randomly located sensor nodes.

Although recent statistical methods are able to detect the boundary nodes, these approaches are only suitable for WSNs with dense or random node distribution. Therefore, these approaches could not be applied to the WSNs with low-density sensor node distribution.

2.3 Topological methods

When topological methods are used for detection of coverage holes, it is not necessary to know the sensor's exact geographic information. Instead, the distance and connectivity information between different nodes are used for hole detection [55].

In [20], Kanno et al. use a hole-equivalent planar graph to preserve hole position. Next a planar simplicial complex called maximal simplicial complex is constructed, which contains the hole information. The approach proposed by Kanno et al. can be used for hole detection in wireless sensor network with or without node position information. In [39], Ramazani et al. proposed a centralized coverage hole detection algorithm. In their approach, a coverage planar graph of the wireless sensor network was constructed, and the locations of the nodes were estimated using received signal strength. Next, simplicial complex was adopted to identify hole boundaries from the coverage planar graph. In [10], De Silva et al. proposed a centralized hole detection algorithm based on the homology of a simplicial pair called Vietoris–Rips complexes. The Rips complex was used to check the coverage by verifying the first homology group of the Rips complex. However, the algorithm proposed in [10] could not detect triangular holes effectively.

In [25], Kröller et al. proposed a combinatorial approach. The approach has two stages. The first stage is hole boundary recognition. The second stage is topology extraction. The proposed approach is based on the assumption that the distribution of nodes is sufficiently dense. In [45], a distributed approach was proposed. In this approach, homology computation of the Rips complex and the combinatorial Laplacian flows were used to verify the existence of coverage holes and obtain the hole position. However, this approach cannot find hole boundary accurately. In [54], Yan et al. proposed a homology-based distributed hole detection method. The proposed method could obtain the location of the holes and effectively detect the nontriangular holes. However, this method could not detect triangular holes effectively, and the complexity was high. In [53], Yan et al. proposed a simplicial complex reduction algorithm for hole detection. Although the detection accuracy was high for non-triangular holes, this algorithm could not detect triangular holes accurately.

Therefore, to overcome the shortcomings of the existing approaches, we propose a novel coverage hole detection method called FD-TL in this paper. The proposed approach is designed to detect triangular and nontriangular holes effectively. In our approach, we use FD algorithms to generate the potential (estimated) layouts of wireless sensor network. Next, we exploit the image recognition capabilities of CNNs to detect coverage holes. Since FD-TL is based on estimated layouts, our method does not depend on the location information generated by GPS. Specifically, the proposed approach is entirely based on network topology information.

2.4 Force-directed algorithms

Force-directed (FD) algorithms have been developed over the last 50 years and applied in many application fields such as information visualization, sensor networks [59][49][12], graph drawing, routing algorithms, scheduling, etc [7]. In addition, FD algorithms are exible, intuitive, and can be easily implemented. They can produce relatively good quality graphs from the network topology alone without the information of nodes' physical location [5]. Due to these reasons, we have adopted FD algorithms for solving hole detection problem in wireless sensor networks.

There are several models of FD algorithms, such as accumulated force models, energy function minimization models, and combinatorial optimization models [8]. Accumulated force models follow the simulation of a spring system. Among the accumulated forces models, there are Eades algorithm [11], Fruchterman-Reingold (FR) algorithm [15] and ForceAtlas2 (FA2) algorithm [18]. Energy function minimisation models use the spring system to minimise the difference between the visual distance and theoretical graphed distance. There is energy function minimization model like Kamada-Kawai (KK) algorithm [19]. Combinatorial optimisation models are probabilistic algorithms, and often inspired by evolutionary mechanisms. Among the combinatorial optimization models, there are Davidson-Harel (DH) algorithm [9] and Kudelka algorithm [26]. In our previous work, we adopted several FD algorithms for evaluation, such as ForceAtlas2 (FA2) algorithm, Kamada Kawai with multiple node selection and decaying stiffness (KK-MS-DS) algorithm [5] and Davidson Harel (DH) algorithm. Because the quality of WSN layout generated by KK-MS-DS algorithm is better than the layout generated by the FA2 and DH algorithms in WSNs, we adopt KK-MS-DS algorithm in the FD-TL method.

The objective of KK-MS-DS algorithm is to push the nodes in the outer boundary away from the inner nodes with a decaying stiffness ($m$) assigned to nodes. The higher the decaying stiffness value ($m$) of a node has, the further the distance of the node can be moved. The KK-MS-DS algorithm includes the following four steps.

Firstly, a node with the highest average degree is selected as the starting node $s$. Two-hop neighbors of starting point $s$ are then selected in order to construct an initial starting area. Secondly, the value of $m$ of the nodes from the starting area decreases along with the execution time by equation (1).

$$m' = m - zp^t \tag{1}$$

where $t$ is the number of times the node has been selected for updating. $p$ is the decay rate and $z$ is the remaining energy the node has.

Thirdly, outside nodes are inserted into the starting area if the the stable status ($r$) of the starting area does not improve for a predefined number of iterations, which is shown as equation (2).

$$r = \frac{\frac{1}{l}\sum_{i=1}^{l}|\hat{L}_i - L_i|}{\sigma} \tag{2}$$

where $l$ is the total number of edges. $\hat{L}_i$ is the edge length on the intermediate iteration and $L_i$ is the edge length on the input network topology. $\sigma$ is the difference in the estimated distance on the input network topology and an intermediate iteration.

The KK-MS-DS algorithm terminates when $r$ remains unchanged up to predefined iterations or $r$ is less than the threshold $\varepsilon$. The pseudo code for the KK-MS-DS algorithm is given in Algorithm 1.

**ALGORITHM 1:** Pseudo code of KK-MS-DS algorithm.

**Input:** Network topology $G = (V, E)$.
**Output:** A boundary ready network topology of $G$.
initialize the start area $WT = (W_{WT}, E_{WT})$ that $WT \subseteq T$;
initialize the iteration count $it = 0$;
initialize the starting node $s$ which has a maximum average degree in $G$;

```
// Step 1.
```
add node $s$ into $V_{WT}$;
**for** $v \in V$ **do**
    **if** $hopcount(V_{WT}, v) \leq 2$ **then**
        add node $v$ into $V_{WT}$;
        $v_{stiffness} = v_{decayingstiffness}$;
    **end**
**end**

```
// Step 2 & 3.
```
**while** $WT = T$ **do**
    $it = it + 1$;
    compute $r$ for the $WT$ every 100 iterations;
    **if** $r < \varepsilon$ **then**
        // The starting area is stable.
        **for** $v \notin V_{WT}, v \in V$ **do**
            **if** $hopcount(V_{WT}, v) \leq 2$ **then**
                add $v$ into $V_{WT}$;
                $V_{stiffness}$ = largest stiffness from its neighbours;
            **end**
            **else**
                $V_{stiffness} = \frac{K}{d_{i,j}^2}$ ;
            **end**
        **end**
        **for** $u, v \in V_{WT}, u \mathrel{/}= v$ **do**
            update $\hat{L}_i$ for node $v$ and $u$;
        **end**
    **else**
        // The starting area is not stable.
        $G = KK - MS(WT, 5\%)$;
        **for** *each* $v \in V_{WT}$ **do**
            $V_{decayingstiffness} = M'$
        **end**
    **end**
**end**

```
// Step 4.
```
clear $V_{WT}$ and $E_{WT}$ in $WT$;
**for** $v \in V$ **do**
    add node $v$ into $V_{WT}$;
    $v_{stiffness} = \frac{K}{d_{i,j}^2}$ ;
**end**
compute $r$ for the $WT$;
**while** $r \geq \varepsilon$ **do**
    compute $r$ for the $WT$ every 100 iterations;
    $G = KK - MS(WT, 5\%)$;
**end**

2.5 Convolutional Neural Networks

CNN is a kind of deep neural network from the area of Artificial Intelligence. CNN was first developed in the 1980s and gained attention in the field of image processing and computer vision. CNNs have been successfully used in various applications such as object detection, semantic segmentation and instance segmentation [27][32][48]. CNNs can detect and recognize general patterns of object from the input image [38]. Particularly, CNNs can be used for object localization of remote sensing images, fault diagnosis and classification in wireless sensor networks.

In [33], Long et al. adopted CNNs to detect objects and solve the problem of object localization in remote sensing images. In [17], Hou et al. made fault diagnosis of rolling bearing based on CNNs and WSN, which improved the reliability of the machinery and operating efficiency. In [47], Tong et al. proposed a CNN-based approach to improve the accuracy of event classification in homogeneous sensor networks. In the proposed FD-TL, CNNs are adopted for coverage hole detection from WSNs. In the hole detection process, CNNs are used to detect both holes' positions and boundaries from the WSNs. In this paper, five CNN models namely Mask R-CNN [16], Mask R-CNN with FPN [29], TensorMask [4], PointRend [23] and BlendMask [3] are adopted for hole detection. In the following paragraphs, we briefly review each CNN models used in the FD-TL.

*Mask R-CNN:* Mask R-CNN [16] is a kind of two-stage framework. Firstly, it scans the image and extracts the feature maps by backbone network. Region Proposal Network (RPN) which is a lightweight neural network looks for area where there is a target and generates region proposals [40]. After getting the final region proposals, it passes the final region proposals to a Region of Interest Align (RoIAlign) layer. Secondly, it makes classification and boundary box regression to obtain the precise position of the bounding box and generate the mask [16]. Compared with Faster R-CNN [40], Mask R-CNN replaces RoIPooling with RoIAlign layer. It also adds a parallel branch network, which outputs a binary mask. RoI Pooling is not pixel-to-pixel alignment, and it leads to inaccurate result in mask generation. To solve the problem, RoIAlign uses bilinear interpolation to find more precise region [16], which could get a better match between feature map and the original image. With RoIAlign layer, the feature map can be matched by pixel of the original image, and the mask precision could be significantly improved. The binary mask is a matrix. The pixels that belong to the target object are represented by 1 and the others are represented by 0. The binary mask indicates whether a given pixel is part of the target object, which could make more accurate object segmentation.

*FPN:* In the object detection task, CNN cannot detect the small object effectively. This is because most of the algorithms only use neural network's high-level feature maps to detect the object. Although the high-level feature maps contains rich global information, it contains little local information which is often important for small object detection. Feature Pyramid Network (FPN) [29] solves the problem of multiple-size object detection. FPN consists of two parts. One is a bottom-up process and another one is the fusion of top-down and lateral connections [29]. The bottom-up process is the same as regular CNN. A CNN can be divided into different layers. Each layer corresponds to one level of the feature pyramid. The top-down process amplifies the high-level feature map to the same size as the previous layer by up-sampling. With up-sampling of the high-level feature map, it could not only make use of the abstract semantic features of the high layer for classification, but also the high-resolution information of the lower layer for location. A lateral connection structure is proposed in [29] to fuse low-level local features and high-level semantic features. Lateral connection fuses the high-level features with the same-resolution features of the previous layer after the high-level features have been up-sampled. By taking full advantage of high-level and low-level features, rich semantic information could be kept while maintaining high resolution in the same time, which greatly improves the effect of small object detection. Aiming to identify small coverage holes effectively, we combine Mask R-CNN [16] with FPN together, and use FPN as the backbone of Mask R-CNN for feature extraction in FD-TL.

*TensorMask:* TensorMask [4] is a kind of dense sliding-window object segmentation algorithm. TensorMask uses structured high-dimensional tensors to represent the image in a set of dense sliding windows [4]. One head of TensorMask is to predict mask, which is responsible for generating mask in sliding window, and the other one is to make classification, which is responsible for predicting target category. In TensorMask, each spatial position's output has its own spatial dimension, which is essentially different from Mask R-CNN. When TensorMask is tested with the Microsoft COCO data set [30], the results show that the performance of TensorMask and Mask R-CNN are similar.

*PointRend:* The aim of PointRend [23] is to optimize the sampling method and predict object's edge effectively. It is similar to the idea of computer graphics rendering. PointRend treats the segmentation problem as a similar rendering problem. Non-uniform sampling is conducted, and the segmentation result of each sampling point is calculated. The main idea is to render the image efficiently by computing only those points that are most likely to differ from the surrounding pixels. Then the result is mapped to a regular grid. The flexible and adaptive point selection strategy is used to predict the segmentation labels [23].

*BlendMask:* The overall architecture of BlendMask[3] contains a detector module and a BlendMask module. The detector module is FCOS [46]. The BlendMask module is composed of three parts. The first part is bottom module, which is used to process the feature maps extracted by backbone network to predict bases. Next the convolution layer which is on detection towers predicts the top-level attention masks and bounding boxes [3]. Finally, blender module fuses bases and corresponding attentions to get the final prediction [3]. The performance of BlendMask is better than Mask R-CNN since BlendMask uses features with higher resolution and better information fusion. BlendMask can encode two kinds of information. One is semantic masks which can judge whether a pixel belongs to an object. The other one is position-sensitive features which can judge whether a pixel is on object's certain part [3]. BlendMask can learn richer feature representation in this way. The base of BlendMask module extracts position-sensitive features which could help to separate instances effectively. Semantic masks could also help to make more precise predictions. As a result, BlendMask performs better in the segmentation of instances. BlendMask fuses the advantage of different models. When testing with Microsoft COCO data set [30], the precision and speed of BlendMask exceed other single-stage instance segmentation algorithms.

## 3 FD-TL Method

In this section, we describe the proposed FD-TL method in details.

### 3.1 Preliminaries

We outline the characteristics of the sensors, connections, and the type of holes considered in our approach as follows:

- We assume that each sensor in the WSN has a unique identification number (node ID).
- The position of each sensor is random and their exact *x* and *y* coordinates are unknown.
- The sensing range ($R_S$) of sensors considered in our approach is uniform and can be defined by users.
- A hole is a polygonal shape consisting of nodes and edges. The interior of the hole can not contain intersecting edges.
- A hole in a WSN must be a closed area and it must be surrounded by sensors.
- In FD-TL, the external holes are not considered since they are actually the boundaries of the WSN.

A simplified example of coverage holes in a WSN is shown as Figure 2. In Figure 2(a), the nodes of the WSN are depicted in pink dots. Next, by using the sensing range of sensor ($R_S$) as the radius, we can draw pink circles around each sensor. These circles are depicted in Figure 2(b). In Figure 2(b), we can observe that some of the areas are not covered in pink colour. These areas represent holes in the WSNs. Finally, based on this observation, we coloured the detected coverage hole in Figure 2(c) using a different colour, which consists of edges surrounding the hole and corresponding boundary nodes.

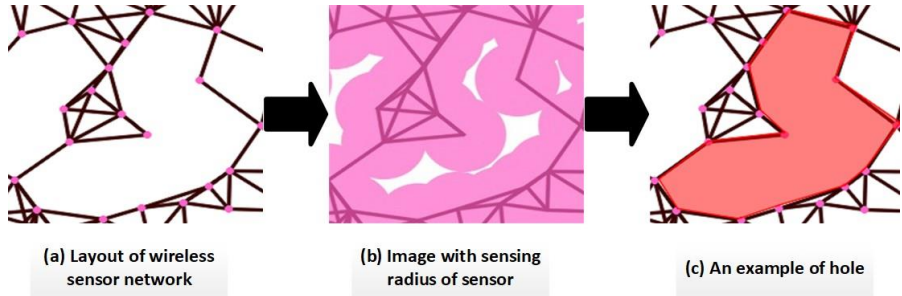

Fig. 2: An example of coverage holes identification based on sensing range ($R_S$).

FD-TL is implemented based on the force-directed (FD) algorithms and Convolutional Neural Network (CNN)s. The overall process of FD-TL can be described in 3 phases; they are model training, testing, and validation processes. The overview of FD-TL is shown in Figure 3. First, in the training phase, we generate training images by a prototype WSN layout generator called CNCAH [6] which was developed based on several state-of-the-art Force-directed Algorithms. In the testing phase, we generate the testing datasets using the same FD algorithms from the CNCAH generator. Next, coverage holes are detected from the testing datasets using trained CNN models. In the validation phase, we compare the output image of CNN models with ground-truth image and evaluate the detection results.

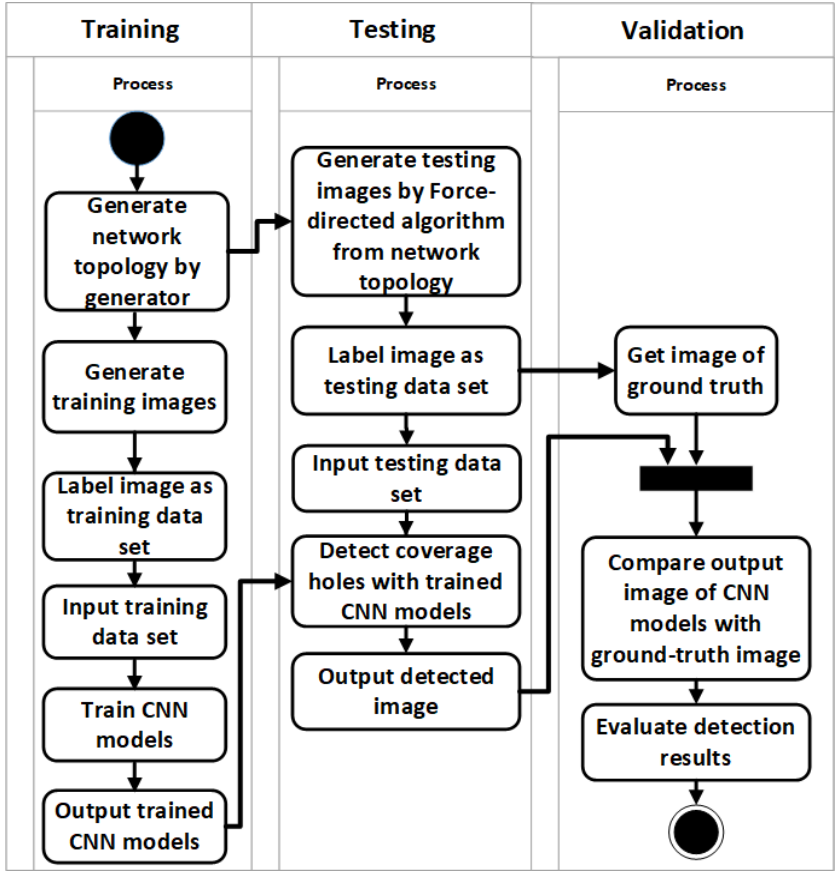

Fig. 3: Overview of FD-TL.

3.2 Training phase

One of the most important requirements of training with CNN models is to prepare appropriate datasets for feature extraction. In the following sections, we detail the process of data generation for CNN models.

*3.2.1 Training data set and validation data set preparation*

To the best of authors' knowledge, there is no WSN coverage hole dataset available in the public domain for training CNN models. Therefore, we prepare a labeled dataset for training the CNN model based on following steps. First, we generate synthetic network topologies consisting of randomly distributed nodes and corresponding edges by a prototype system called CNCAH network generator [6][1].

Force-directed (FD) algorithm is a kind of graph drawing algorithm. FD algorithms can generate visualizations based on information contained in the network topology which only contains a list of nodes and the edges representing these nodes. FD algorithm's output is the approximate layout of the wireless sensor network. In other words, a layout could be considered as one of many possible snapshots of the WSN. Due to the variation in force calculation, for a given input (topology), different layouts will be generated by each FD algorithm at a specific point in time. Moreover, in majority of the FD algorithms, the layouts generated from the input topology can be exported into image or text format for post-processing. In our work, we adopt KK-MS-DS algorithm for evaluation. The training and validation data set preparation process is depicted in Figure 4.

---
[1] https://www.cis.um.edu.mo/ fstasp/tool eric CNCAHNetGenerator.html

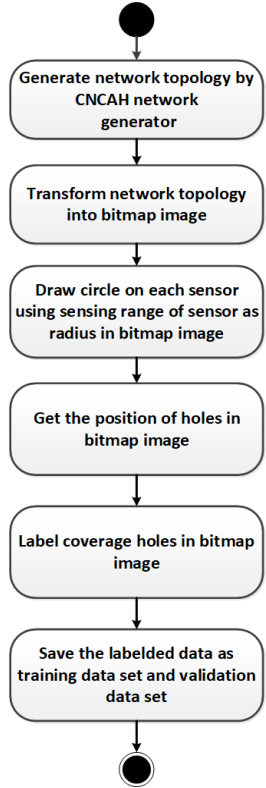

Fig. 4: Training data set and validation data set preparation process.

During the process, we transform each generated network topology into a bitmap image (1024 x 1024). Next, we draw circle on each sensor using sensing range of sensor as radius in bitmap image. The white areas which are not covered by pink circle represent holes. Based on this process, we could identify the position of holes in the bitmap image. The example of a network topology, corresponding bitmap image, and the bitmap image with sensing range of sensor are depicted in Figure 5.

By using the images generated from the above process, we label the coverage holes in the bitmap images manually with LabelMe tool [42] according to the definitions stated in Section 3.1. The image labelling process is shown as Figure 6. The image labelling process has 3 steps. First, we input the bitmap images into the LabelMe tool. Second, we label the coverage holes manually with the LabelMe tool. Third, the labeled image and annotation file are recorded with the nodes' coordinates which are located on the boundaries of the coverage holes. Finally, we split the labeled data randomly into training and validation dataset.

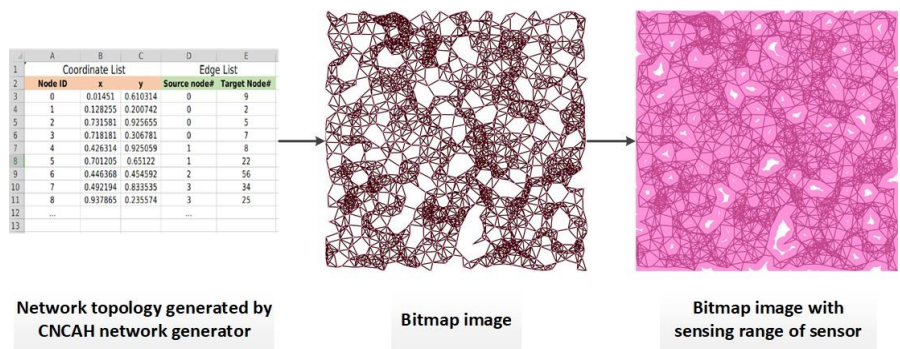

Fig. 5: The network topology and corresponding bitmap image.

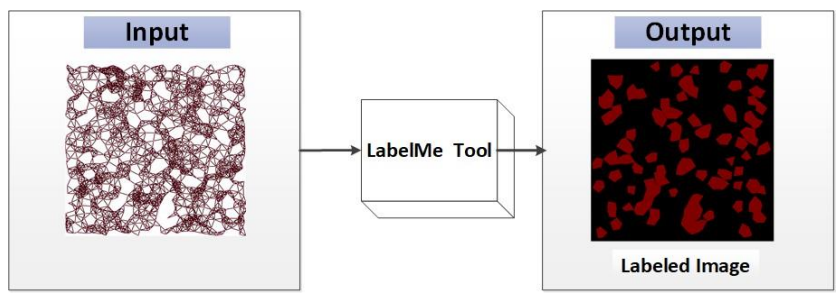

Fig. 6: The image labelling process.

*3.2.2 Training CNN models with transfer learning method*

There are many successful applications of CNN in computer vision such as image classification, localization and detection [22]. CNN is mainly composed of convolutional layers, pooling layers, and fully connected layers [21]. Convolution layer and pooling layer connect to each other alternately in the network. Convolution layer extracts the image's features and outputs the feature map as input into the next layer. Pooling layer performs down-sampling to reduce the training parameters. Pooling layer is also used to avoid over-fitting and to make sure that the object's global information is extracted in the final convolutional layer. The function of fully connected layer is to fuse the extracted feature map and to make the prediction.

In the hole detection process, CNN detects both the positions of holes and the boundary of the holes. It means that CNN needs to effectively recognize the holes through the bounding boxes and locate the precise pixels of each hole to achieve pixel-level segmentation. Therefore, this paper uses Mask R-CNN [16], Mask R-CNN with FPN [29], TensorMask [4], PointRend [23] and BlendMask [3] respectively in FD-TL method to detect coverage holes.

Sufficient training datasets and effective feature extraction method are essential for designing a CNN model which contains a large number of parameters. Without sufficient datasets for training, the generalization ability of CNN could be significantly affected and result in over-fitting. In this situation, the model can be affected by irrelevant information such as noise from the training data. During the training period, it is also difficult for the loss function to converge. Besides, CNN model cannot learn effective features and information if there is insufficient training data. To alleviate the problem of insufficient training data, researchers have proposed an approach to make use of pretrained CNN models that have been trained on large amount of public data sets, and apply the structures and weights to solve the specific problem. This approach is commonly referred to as transfer learning [37]. In transfer learning, the universal feature representation and low-level features such as contours and textures are learnt by a pretrained model. After the low-level feature has been extracted by the pretrained model, we fine tune the weight of higher layers by training with our own datasets. With transfer learning, a better training result can be obtained even though sufficient training datasets are unavailable. In addition, there are a large number of parameters needed to be trained and updated in CNN models. With transfer learning, we can also reduce the number of parameters needed to be trained and updated.

In our proposed method, Mask R-CNN [16] model, Mask R-CNN with FPN [29] model, TensorMask [4] model, PointRend [23] model and BlendMask [3] model are first pretrained with the Microsoft COCO data set [30]. Next, these models are fine-tuned with our own hole detection training datasets which is are obtained in the previous step. The overall training process is depicted in Figure 7.

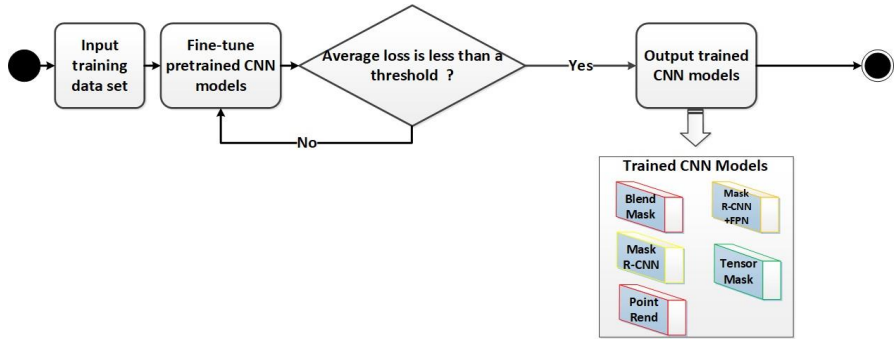

Fig. 7: Training phase of FD-TL method.

3.3 Testing phase

The testing phase of FD-TL is depicted in Figure 8. First, we generate the topology of wireless sensor network by the CNCAH network generator [6]. Next we input the topology of wireless sensor network into KK-MS-DS [19] algorithm which is a kind of Force-directed algorithm. The topology of wireless sensor network contains all sensors' connection information. Then KK-MS-DS algorithm generates the layout of wireless sensor network from the input topology.

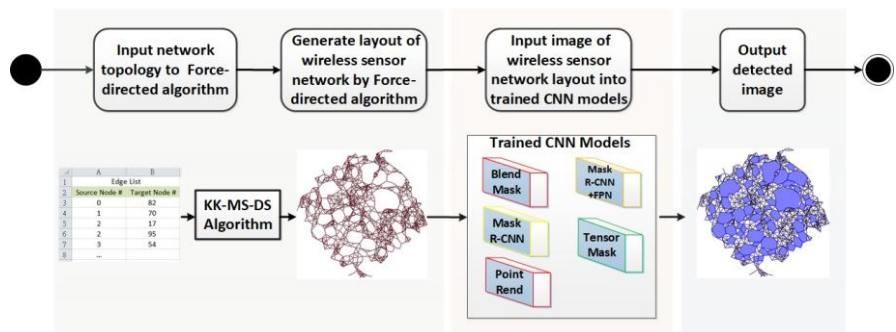

Fig. 8: Testing phase of FD-TL method.

For any input topology of a WSN, the KK-MS-DS algorithm is able to a generate layout at different time points, which approximates the actual physical layout of the WSN. The generated layout is gradually improved as the execution time grows. The KK-MS-DS algorithm adjusts the position of different nodes according to the force calculation, and the iteration process terminates when the nodes' position remain unchanged. Each iteration produces a layout, and we use the layouts as the test dataset for hole detection. The process of iteratively improving the generated layouts by KK-MS-DS algorithm is illustrated in Figure 9(a) and Figure 9(b). The layouts of a WSN shown in Figure 9(a) and Figure 9(b) are generated by the KK-MS-DS algorithm at the $20^{th}$ and $100^{th}$ second respectively. The topology of the WSN contains 500 nodes and 1500 edges.

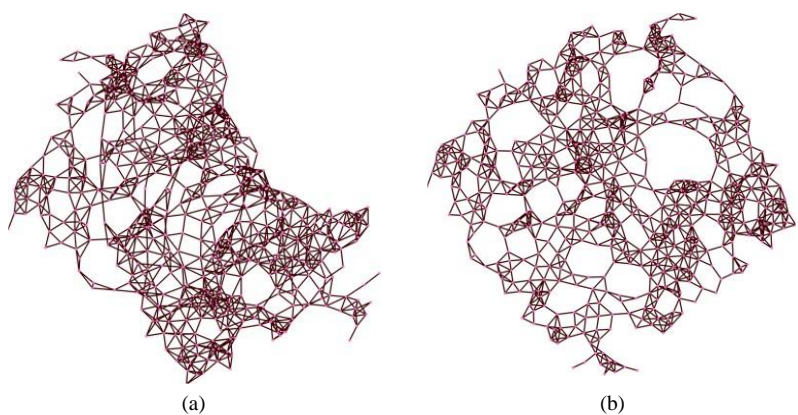

(a)     (b)

Fig. 9: Layout generated by KK-MS-DS algorithm at (a) $20^{th}$ second and (b) $100^{th}$ second.

Next, we input the layouts of a wireless sensor network generated by KK-MS-DS algorithm into the trained CNN model for hole detection. The sample output images which contain holes detected by a CNN model are depicted in Figure 10(a) and Figure 10(b).

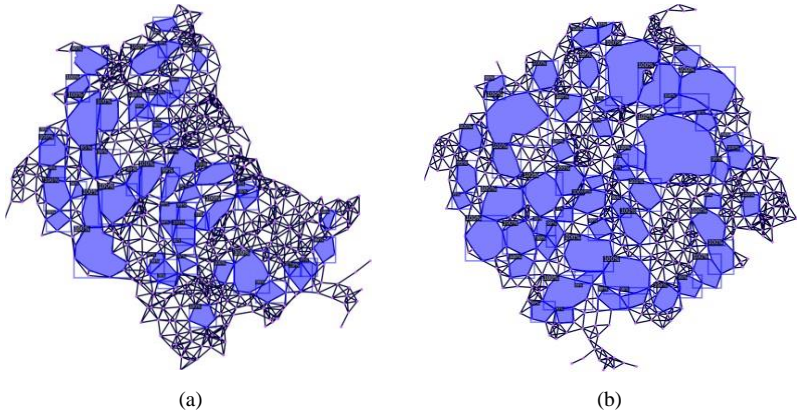

(a)  (b)

Fig. 10: Output images of a CNN model at (a) $20^{th}$ second and (b) $100^{th}$ second.

### 3.4 Validation phase

To evaluate the hole detection results, we compare the holes detected by CNN models with the ground truth from validation dataset generated in section 3.2.1. Recall that FD algorithms are designed to approximate the physical layouts of the input topology using calculated force (energy) value. As the execution progresses, KK-MS-DS algorithm iteratively adjust the locations of the nodes (i.e. sensors). Therefore, for any input topology, the layout of wireless sensor network generated by KK-MS-DS algorithm can vary at different execution time points. Although such adjustment is being carried out at each time point, the node ID and the connection information between different nodes remain the same for a specific input topology. In the validation phase, we identify the node IDs from the boundary of the detected holes. Next, the IDs of these holes are compared against the holes from the ground truth dataset. Pseudo code for identifying the ID of the nodes along the hole's boundary from the ground truth and from the output image of the CNN model is listed in Algorithm 2. The overall procedure of the validation phase is depicted in Figure 11. If the ID of the node along the boundary in a detected hole is the same as that of ground truth, the hole is counted as correctly detected by the trained CNN models.

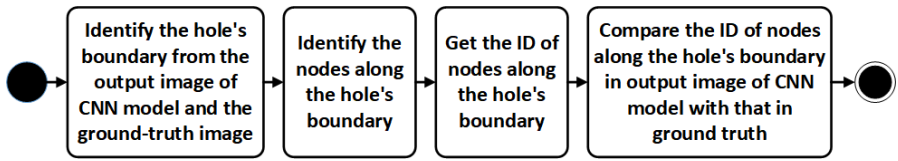

Fig. 11: The validation phase of FD-TL.

**ALGORITHM 2:** Algorithm for identification of node's ID along the holes' boundary.

**Input:** Ground-truth image, the output image of CNN model, network topology of wireless sensor network($T$).
**Output:** The ID of nodes along the hole's boundary($B$-$nodes$ $ID$) in the ground-truth image and the output image of CNN model.

`// Variables(V) initialization.`
let $V$ be the set of nodes in $T$;
let hole color be red in the ground-truth image;
let hole color be blue in the output image of CNN model;
$B - nodes = \{\}$;
$B - nodes - ID = \{\}$;
`// 1) Find the hole contours from the ground-truth image and the output image of CNN model with color.`
let $C$ be the list of contours found from the ground-truth image and the output image of CNN model;
**for** *each contour c in C* **do**
    **for** *each node v in V* **do**
        **if** *PointTest(v, contour) = TRUE* **then**
            `// 2) Save the specific nodes along the hole's boundary.`
            $B - nodes \leftarrow \{v\}$;
        **end**
    **end**
**end**
`// 3) Return the ID of nodes along the hole's boundary.`
**return** $B - nodes - ID$.

*3.4.1 Evaluation criteria*

The notations used in this paper are summarized in Table 1. The performance evaluation criteria are listed in Table 2. Note that we did not measure the precision criteria in the experiments because of the imbalanced data problem. In WSNs, the number of nodes belonging to holes and the number of nodes not belonging to holes are imbalanced. In other words, most of the nodes from the test dataset are not on the boundary of the holes. Therefore, the number of TP and TN in datasets vary significantly. Because of the imbalanced data, the precision of hole detection cannot be used to reflect the CNN models' real performance. Aiming to solve this problem, we use sensitivity and specificity as the benchmark criteira for evaluating the CNN's performance.

Table 1: Notations.

| Notation | Description |
|---|---|
| $n$ | The number of nodes of the wireless sensor network. |
| $d$ | Average degree of the wireless sensor network, $d = 2(num_{edge}/num_{node})$. |
| $TP$ | The node which is on the detected hole's boundary as well as on the actual hole boundary. |
| $TN$ | The node which is not on any of the detected holes' boundary. This node is also not on any of the actual hole boundaries. |
| $FP$ | The node which is on the detected hole's boundary but not on the actual hole boundary. |
| $FN$ | The node which is not on any of the detected holes' boundary. But, this node is on the actual hole boundary. |

Table 2: Performance evaluation criteria.

| Criteria | Description | |
| --- | --- | --- |
| Sensitivity | $Sensitivity = TP/(TP + FN)$ | The higher the better |
| Specificity | $Specificity = TN/(TN + FP)$ | The higher the better |

## 4 Experiments

In the experiments, we evaluate the performance of FD-TL based on the evaluation criteria outlined in the previous section. In the experiments, FD-TL with Mask R-CNN, Mask R-CNN with FPN, TensorMask, PointRend and BlendMask are evaluated based on WSNs different average degree and number of nodes.

### 4.1 Experiment settings

The experiments were conducted on computer with Intel Core i7-10875H processor, graph card NVidia Geforce RTX2070S with 16GB's memory. The configuration parameters of CNN models are listed in Table 3.

Table 3: The configuration parameters of CNN models.

| Configuration | BlendMask | Mask R-CNN + FPN | Mask R-CNN | TensorMask | PointRend |
| --- | --- | --- | --- | --- | --- |
| Activation Function | RELU | RELU | RELU | RELU | RELU |
| Normalization | Batch Normalization, Group Normalization | Batch Normalization, Group Normalization | Batch Normalization, Group Normalization | Batch Normalization, Group Normalization | Batch Normalization, Group Normalization |
| Base Learning Rate | 0.01 | 0.01 | 0.01 | 0.01 | 0.01 |
| Warmup Iterations | 1000 | 1000 | 1000 | 1000 | 1000 |
| Weight Decay | 0.0001 | 0.0001 | 0.0001 | 0.0001 | 0.0001 |
| Backbone | ResNet-50-FPN | ResNet-50-FPN | ResNet-50-C4 | ResNet-50-FPN | ResNet-50-FPN |
| Images_Per_Batch | 2 | 2 | 2 | 1 | 2 |
| Input_Image_Size | 1024*1024 | 1024*1024 | 1024*1024 | 1024*1024 | 1024*1024 |

In the experiments, the number of node $n$ is set to 500, 1,000, and 2,000. The average degree $d$ is set to 6, 8, and 10 respectively to generate the topologies. In total, 9 topologies of wireless sensor network are generated by using CNCAH network generator [6]. Next KK-MS-DS algorithm is used to generate the layouts of wireless sensor network from the 9 topologies. During the execution [19], KK-MS-DS algorithm improves and updates the layout of wireless sensor network iteratively, and the iteration process terminates when the nodes' position remain unchanged. Each iteration produces a layout, and we use the layout as the test data set for hole detection with trained CNN models. Figure 12 shows an example of a wireless sensor network with $n = 500$ and $d = 6$, where the pink dots represent the location of each sensor.

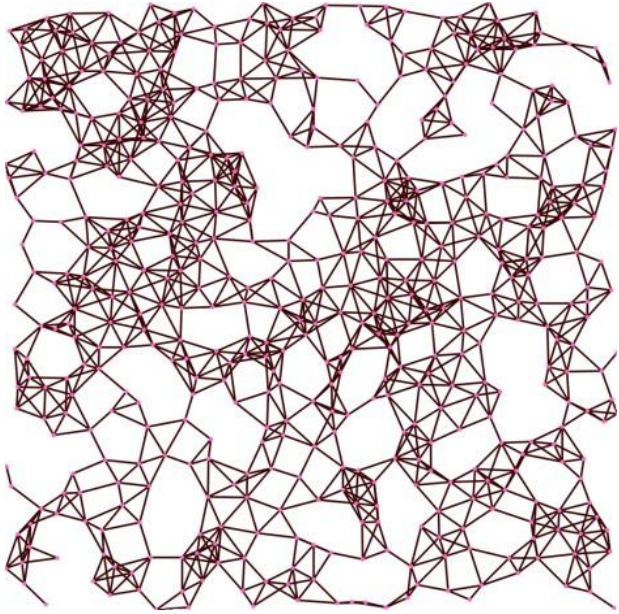

Fig. 12: An example of wireless sensor network with $n = 500$ and $d = 6$.

4.2 Experiment results of wireless sensor networks with 500 nodes

In this experiment, the number of node $n$ is set to 500, and the average degree $d$ is set to 6, 8, and 10 respectively. The experimental results from Figure 13 show that FD-TL can obtain 85% sensitivity in all cases. Specifically, the sensitivity can reach to 90% when the average degree is 6 and 10 (see Figure 13(a) and Figure 13(c)). We can also observe that all CNN models can achieve relatively stable sensitivity when they detect coverage holes.

Experiment results in Figure 14 show that FD-TL achieves at least 89% specificity for 5 different CNN models when $n = 500$. Specifically, when degree is 6, the specificity of FD-TL is similar for 5 different CNN models, which is above 95% (see Figure 14(a)). As shown in Figure 14(c), when degree is 10, BlendMask performs better than TensorMask and PointRend, and the specificity of Mask R-CNN with FPN structure is the same as that of Mask R-CNN without FPN structure, which performs the worst. According to the experimental results depicted in Figure 14(a) and Figure 14(c), the specificity of FD-TL with a low average degree (d = 6) is better than that with a high average degree (d = 10).

In summary, all CNN models show a stable performance for coverage hole detection when $n = 500$. Sensitivity and specificity for different average degrees are above 85%. When average degree is 6, sensitivity and specificity of FD-TL are above 90% and 96% respectively.

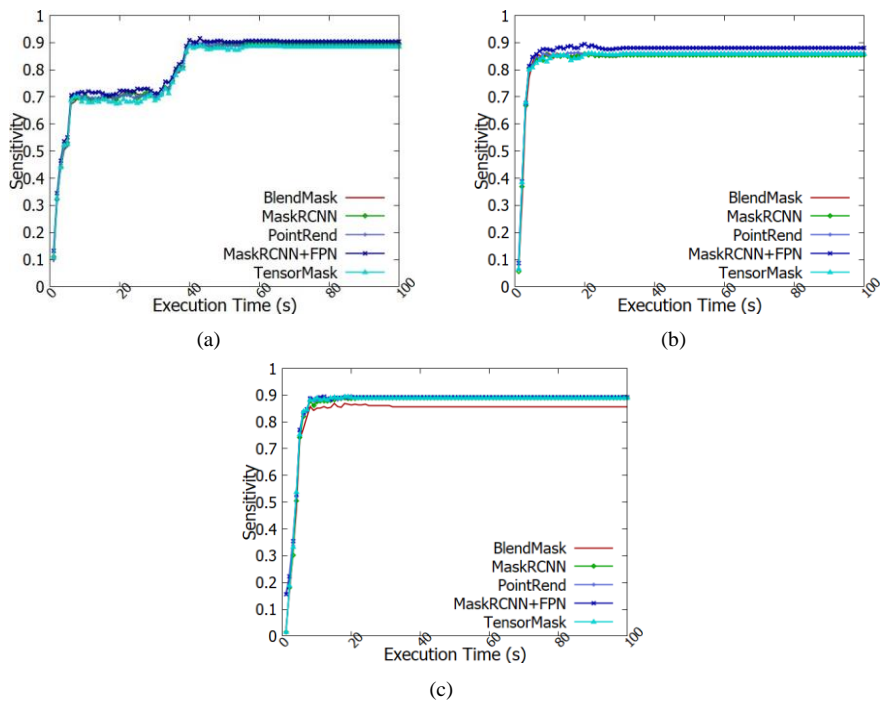

Fig. 13: Sensitivity of hole detection when $n = 500$ and (a) $d = 6$, (b) $d = 8$ and (c) $d = 10$ in wireless sensor networks.

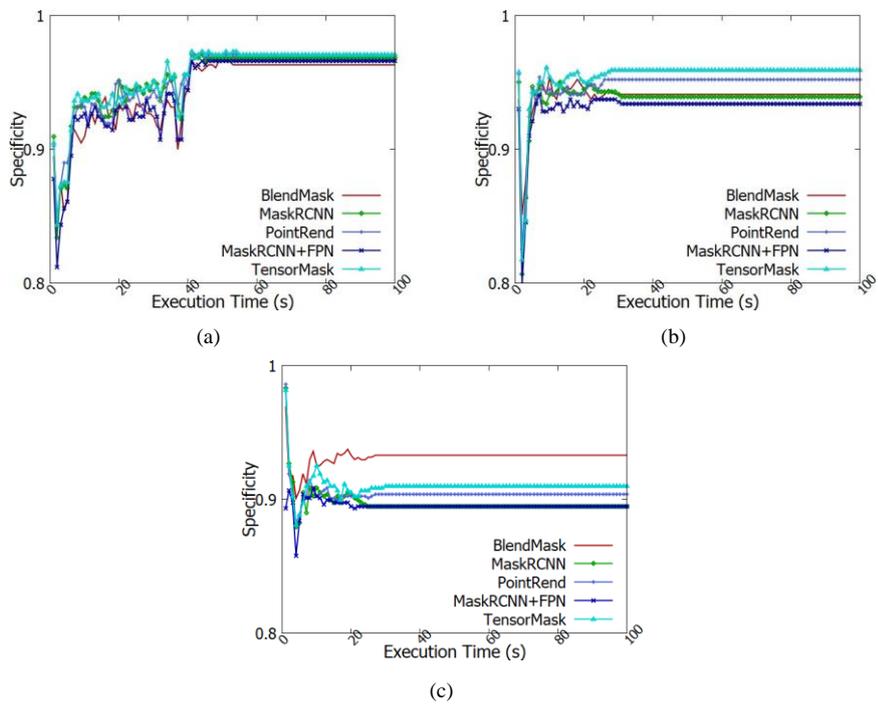

Fig. 14: Specificity of hole detection when $n = 500$ and (a) $d = 6$, (b) $d = 8$ and (c) $d = 10$ in wireless sensor networks.

4.3 Experiment results of wireless sensor networks with 1000 nodes

In this experiment, the number of node $n$ is set to 1000, and the average degree $d$ is set to 6, 8, and 10 respectively. The experimental results in Figure 15 show that FD-TL can achieve approximately 80% sensitivity for all CNN models. Specifically, the sensitivity can reach to 90% when the average degree is 8 (see Figure 15(b)), which is higher than the sensitivity when average degree is 6 and 10. According to the experimental results depicted in Figure 15(a) and Figure 15(c), the sensitivity of FD-TL with a low average degree ($d = 6$) is better than that with a high average degree ($d = 10$).

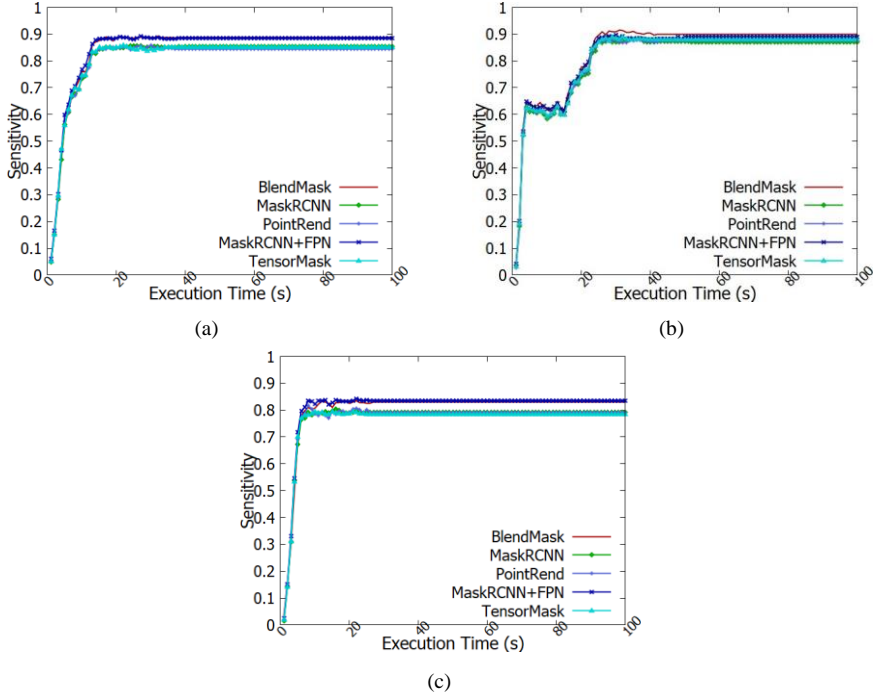

Fig. 15: Sensitivity of hole detection with $n = 1000$ and (a) $d = 6$, (b) $d = 8$ and (c) $d = 10$ in wireless sensor networks.

The experimental results in Figure 16 show that FD-TL can achieve approximately 84% specificity in majority of the cases when $n = 1000$. The specificity of FD-TL with a high average degree ($d = 10$) is better than that with a low average degree ($d = 6$ and 8). When the average degree is 10, FD-TL can obtain above 90% specificity for 5 different CNN models (see Figure 16(c)). As shown in Figure 16(a), when degree is 6, the performance of TensorMask, PointRend and Mask R-CNN is similar, and the specificity of BlendMask is approximately equal to that of Mask R-CNN with FPN structure.

When $n = 1000$, the sensitivity and specificity for different average degrees are above 80% and 84% respectively. When average degree is 8, the sensitivity and specificity of FD-TL could reach to 90%.

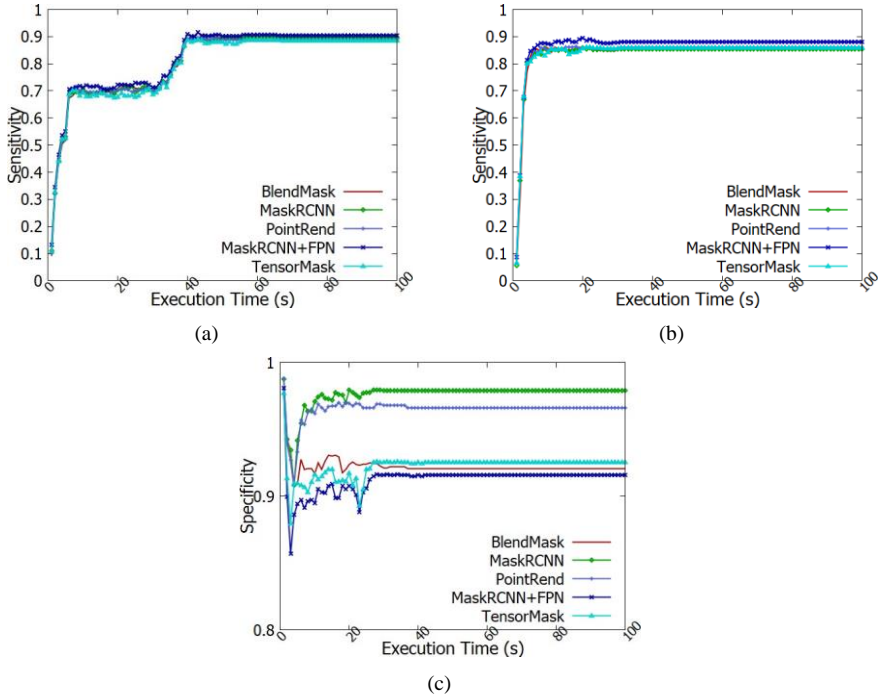

(a)                                                                                       (b)

(c)

Fig. 16: Specificity of hole detection with *n* = 1000 and (a) *d* = 6, (b) *d* = 8 and (c) *d* = 10 in wireless sensor networks.

4.4 Experiment results of wireless sensor networks with 2000 nodes

In this experiment, number of nodes *n* is set to 2000, and the average degree *d* is set to 6, 8, and 10 respectively.

*4.4.1 Sensitivity*

The experimental results in Figure 17 reveal that five CNN models used in the experiment have varying performance for different average degree. From Figure 17, we can observe that BlendMask algorithm achieves the best performance among all CNN algorithms. Specifically, the sensitivity obtained by BlendMask algorithm is above 90% for all of average degrees when *n* = 2000. It is because BlendMask can learn a richer feature representation by extracting position-sensitive features. Besides, BlendMask is able to perform better low-layer and high-layer information fusion. Since position-sensitive features are sensitive to different positions, BlendMask could judge whether a specific pixel is in a certain part of the coverage hole or not. This capability can help to separate holes better and as a result, the segmentation and performance in detection is significantly improved.

When Mask R-CNN uses FPN as its backbone for feature extraction network, its performance is better than Mask R-CNN without FPN structure. On average, the sensitivity of Mask R-CNN with FPN structure is 10% higher than the sensitivity of Mask R-CNN without FPN structure when degree is 6, 8 and 10. Specifically, sensitivity of Mask R-CNN with FPN structure is 89% when degree is 8, while the sensitivity of Mask R-CNN without FPN structure is 67% (see Figure 17(b)). The reason behind the difference in performance is caused by that fact that the FPN structure uses a multi-level feature map to detect coverage holes of different sizes. The experiment results show that the FPN structure is indeed effective in detecting coverage holes from test datasets. From the experiment results, we can also observe that the sensitivities of TensorMask algorithm when $d = 6$, 8 and 10 are lower than 75% (see Figure 17(a), (b), and (c)). According to the results from Figure 17(a) and Figure 17(b), the performance of PointRend algorithm is the worst among all CNN algorithms tested. Compared with the previous experiments on $n = 500$ and 1000, the sensitivity of the PointRend algorithm drops sharply when $n = 2000$. It shows that the performance of PointRend is unstable. The overall performance of PointRend is poor when when $n$ is set to 2000. All in all, we can observe that the sensitivity of coverage hole detection for WSNs with $n = 2000$ is less promising than WSNs with $n = 500$ and 1000. It is because sensor distribution is denser when $n = 2000$. As a result, some CNN models such as Mask R-CNN, TensorMask and PointRend algorithm cannot remain stable performance. Among the five CNN models, BlendMask performs the best, whose sensitivity is approximately 90% when $n$ is 500, 1000 and 2000. PointRend performs the worst, whose sensitivity drops to 55% when $n$ is 2000.

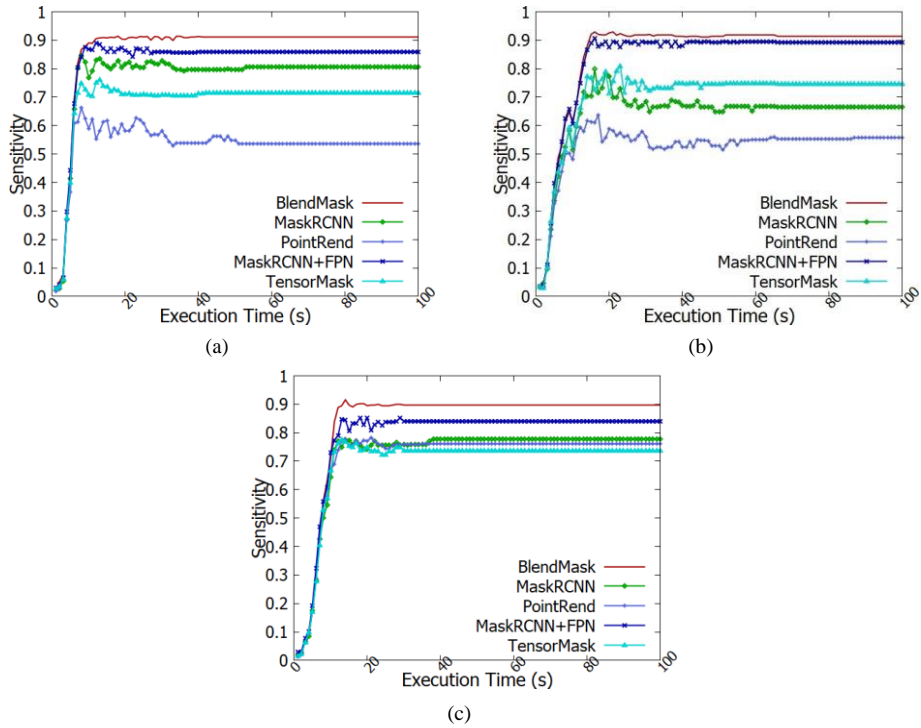

Fig. 17: Sensitivity of hole detection with $n = 2000$ and (a) $d = 6$, (b) $d = 8$ and (c) $d = 10$ in wireless sensor networks.

*4.4.2 Specificity*

The experimental results illustrated in Figure 18 show that the specificity of different degrees present very different performance with 5 CNN models when $n = 2000$. Among all of CNN models, PointRend has the highest specificity, but its sensitivity is very low. PointRend algorithm cannot correctly detect the coverage hole when $n = 2000$, and it tends to predict the true hole as non-hole, which resulting in low sensitivity and high specificity. The performance of TensorMask algorithm and Mask R-CNN algorithm are similar when $n = 2000$, whose specificity could be above 90% when degree is 8 and 10 illustrated as Figure 18(b) and Figure 18(c). Specificity of Mask R-CNN with FPN algorithm is high when average degree is 8 and 10, but its specificity is relatively low when average degree is 6. As shown in Figure 18(c), when average degree is 10, the specificity of BlendMask can reach to 90%. However, when average degree is 6 and 8, the specificity are low, which indicates that BlendMask tend to predict true non-hole as hole when $d = 6$ and 8, and it results in high sensitivity and low specificity.

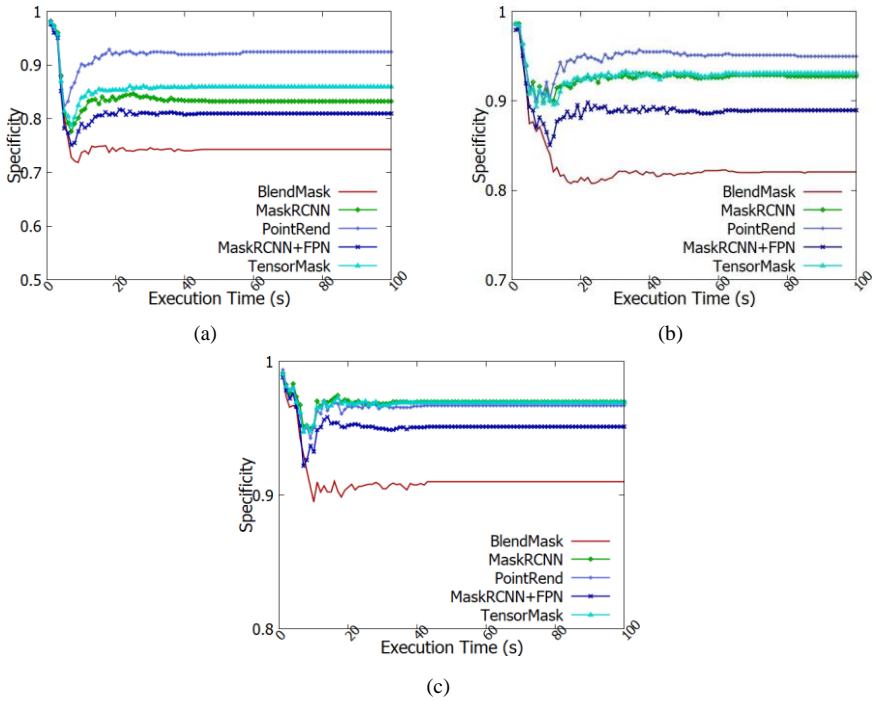

Fig. 18: Specificity of hole detection with $n = 2000$ and (a) $d = 6$, (b) $d = 8$ and (c) $d = 10$ in wireless sensor networks.

Based on the analysis, when $n = 2000$, the sensitivity and specificity of 5 CNN models all show huge difference with $d = 6$, 8 and 10. When $n = 2000$, some CNN models such as PointRend, TensorMask and Mask R-CNN cannot detect coverage holes precisely. The CNN model with the best performance is BlendMask and it could detect holes efficiently.

In summary, we analyze the results of coverage hole detection using different CNN models such as Mask R-CNN, Mask R-CNN with FPN, TensorMask, PointRend and BlendMask in the experiments. Our analysis also reveals the reasons behind the different performance results of CNN models. According to the results, when $n = 500, 1000$ and $2000$, BlendMask achieves the best result because it can learn a richer feature representation by extracting position-sensitive features. BlendMask also makes better low-layer and high-layer information fusion, which allows the BlendMask to accurately detect and segment the coverage holes of different positions. The performance of Mask R-CNN with FPN structure is better than that of Mask R-CNN without FPN structure, which proves the effectiveness of FPN structure in coverage hole detection. Mask R-CNN without FPN structure and TensorMask have relatively low sensitivity when $n = 2000$, which is worse than BlendMask. Their sensitivity decreases when $n$ increases from 500 to 2000, which means that their performance is not stable enough for detecting holes. We found that the performance of PointRend is unstable. PointRend cannot detect holes effectively when the number of node is set to 2000. The experimental results show that the non-uniform sampling and adaptive point selection method of PointRend did not perform well for the hole detection task when $n$ reaches to 2000. From the experiment results, we found that the performance of CNN models for coverage hole detection deteriorates when the number of nodes $n$ is increased. It is because when $n$ is increased, sensor distribution becomes denser.

## 5 Conclusions

In this paper, we propose a novel hole detection approach for wireless sensor network called FD-TL which is based on Force-directed Algorithms and Convolutional Neural Networks (CNN) with transfer learning. To the best of authors' knowledge, the proposed FD-TL is the first hole detection method which is based on FD algorithm and CNN. The experimental results show that the FD-TL method can effectively detect the coverage holes from WSNs. The FD-TL method can achieve 90% sensitivity and 96% specificity for hole detection in wireless sensor networks. In terms of efficiency, the proposed FD-TL method can detect coverage holes from an image within 1 second.

As for the future work, we are planning to optimize FD-TL method for hole detection from wireless sensor networks when the number of nodes is larger than 2000. Besides, we also plan to extend FD-TL method as a cornerstone of future research that guides the FD algorithm to iteratively estimate the layout of wireless sensor networks from input topologies.

**Acknowledgements** This research was funded by University of Macau (File no. MYRG2019-00136-FST and MYRG2017-00029-FST).